\documentclass[aps,twocolumn,nofootinbib,floatfix,superscriptaddress,preprintnumbers]{revtex4-1}

\usepackage{colortbl}

\definecolor{Gray}{gray}{0.95}
\definecolor{RGray}{gray}{0.85}
\definecolor{CGray}{gray}{0.92}

\usepackage{amsmath,amssymb,bm,multirow,float,bbold,array,booktabs,arydshln}
\usepackage{graphicx,color}
\usepackage[usenames,dvipsnames]{xcolor}
\usepackage{dsfont} 
\usepackage[colorlinks,bookmarksopen,bookmarksnumbered,pdfstartview=FitH,citecolor=SBred, linkcolor=SBred,urlcolor=SBred]{hyperref}
\usepackage{graphicx}
\usepackage{ulem} 
\usepackage{enumitem}

\definecolor{SBred}{rgb}{0.6471, 0.1098, 0.1882}

\def\nvl{n_{\text{VL}}}
\def\a23{\alpha_{23}}

\def\EW{\mathrm{SU(2)}_L \otimes \mathrm{U(1)}_Y}

\newcommand{\GeV}{\,{\rm GeV}}
\newcommand{\TeV}{\,{\rm TeV}}

\newcommand{\be}{\begin{equation}}
\newcommand{\ee}{\end{equation}}

\newcommand{\eq}[1]{\begin{equation} #1 \end{equation}}
\newcommand{\eqa}[1]{\begin{eqnarray} #1 \end{eqnarray}}

\newcommand{\ord}{\mathcal{O}}
\renewcommand{\L}{{\cal L}}
\newcommand{\C}{{\cal C}}

\newcommand{\Fig}[1]{Fig.~\ref{#1}}
\newcommand{\Eq}[1]{Eq.~(\ref{#1})}
\newcommand{\Tab}[1]{Table~\ref{#1}}
\newcommand{\Sec}[1]{Section~\ref{#1}}

\newcommand\VRule[1][\arrayrulewidth]{\vrule width #1}

\def\beqn#1{\begin{equation}\label{#1}}
\def\eeqn{\end{equation}}
\def\beqa#1{\begin{eqnarray}\label{#1}}
\def\eeqa{\end{eqnarray}}

\begin{document}

\preprint{SI-HEP-2016-10}
\preprint{QFET-2016-05}
\preprint{LMU-ASC 16/16}
\preprint{IFIC/16-19}

\title{Non-abelian gauge extensions for $\mathbf{B}$-decay anomalies}

\author{Sofiane M. Boucenna}
 \affiliation{INFN, Laboratori Nazionali di Frascati, C.P. 13, 100044 Frascati, Italy}
\author{Alejandro Celis}
 \affiliation{Ludwig-Maximilians-Universit\"at M\"unchen, 
   Fakult\"at f\"ur Physik,\\
   Arnold Sommerfeld Center for Theoretical Physics, 
   80333 M\"unchen, Germany}
\author{Javier Fuentes-Mart\'{\i}n}
\affiliation{Instituto de F\'{\i}sica Corpuscular, Universitat de Val\`encia - CSIC, E-46071 Val\`encia, Spain}
\author{Avelino Vicente}
\affiliation{Instituto de F\'{\i}sica Corpuscular, Universitat de Val\`encia - CSIC, E-46071 Val\`encia, Spain}
\author{Javier Virto}
\affiliation{\mbox{Theoretische Elementarteilchenphysik, Naturwiss.- techn. Fakult\"at,}\\
Universit\"at Siegen, 57068 Siegen, Germany}


\begin{abstract}
\vspace{5mm}

We study the generic features of minimal  gauge extensions of the Standard Model in view
of recent hints of lepton-flavor non-universality in semi-leptonic $b \to s \ell^+ \ell^-$ and $b \to c \ell \nu$ decays. We classify the possible
models according to the symmetry-breaking pattern and the source of flavor
non-universality. We find that in viable models the $\mathrm{SU(2)}_L$ factor is embedded non-trivially in the extended gauge group,
and that gauge couplings should be universal, hinting to the presence of new degrees of freedom sourcing non-universality.
Finally, we provide an explicit model that can explain the $B$-decay anomalies in a coherent way and confront it with the relevant phenomenological constraints.

\vspace{3mm}
\end{abstract}

\maketitle

\section{Introduction}\label{sec:intro}

Low-energy experiments have been crucial in the development of the current Standard Model (SM) of
electroweak interactions, based on the gauge group $\mathrm{SU(2)}_L \otimes \mathrm{U(1)}_Y$.
The structure of the SM electroweak theory was beautifully revealed by a large variety of experimental
observations at low energy, together with requirements of a proper high-energy behavior of the theory.
In particular, the intermediate vector bosons $W^{\pm}, Z$ were predicted theoretically before their
experimental discovery.
Precision experiments at low energies continue providing important information about the possible ultraviolet (UV)
completions of the SM, and New Physics (NP) might be revealed again first through the precision frontier.\\

Currently there are two sets of interesting tensions in $B$-physics data:\\[-2mm]

{\bf 1.}
In 2012 the BaBar collaboration reported deviations from lepton universality at the $25\%$ level
in the exclusive semileptonic $b\to c$ decays, through a measurement of the ratios
\begin{align}
R(D) &= \frac{ \Gamma(B \rightarrow D \tau \nu )}{\Gamma(B \rightarrow D \ell \nu )}  \stackrel{\rm SM}{=}  0.297 \pm 0.017 \,, \\[0.2cm]
R(D^{*}) &= \frac{ \Gamma(B \rightarrow D^{*} \tau \nu )}{\Gamma(B \rightarrow D^{*} \ell \nu )} \stackrel{\rm SM}{=}  0.252 \pm 0.003   \,, 
\end{align}
with $\ell =e~\text{or}~\mu$.
The measured values by BaBar~\cite{Lees:2012xj}, $R(D) = 0.440 \pm 0.072$ and $R(D^*) = 0.332 \pm 0.030$,
show an excess with respect to the SM of  $2.0\,\sigma$ and $2.7\,\sigma$
respectively~\cite{Lees:2012xj,Kamenik:2008tj,Fajfer:2012vx}.
The Belle collaboration reported a measurement of these ratios in 2015 which
showed a slight enhancement with respect to the SM, $R(D) = 0.375 \pm 0.069$ and $R(D^*) = 0.293 \pm 0.041$~\cite{Huschle:2015rga}.
The LHCb collaboration also measured $R(D^*) = 0.336 \pm 0.040$~\cite{Aaij:2015yra}, representing a deviation
from the SM at the $\sim 2\,\sigma$ level.
Very recently, the Belle collaboration has presented a new independent determination of
$R(D^*)$~\cite{Abdesselam:2016cgx} which is $1.6\,\sigma$ above the SM and is compatible with all the previous
measurements: $R(D^*) = 0.302 \pm 0.032$.\\[-2mm]

{\bf 2.} The LHCb collaboration has provided as well hints for flavor non-universality (FNU) in
$b \rightarrow s \ell^+ \ell^-$ transitions.  The ratio~\cite{Hiller:2003js}\footnote
{
We note that electromagnetic corrections to this ratio are expected to be of order
$\alpha \log(m_e^2/m_\mu^2)\sim 8\%$. These logarithmic terms could also be enhanced by non-perturbative
effects of order $\log(\Lambda/m_B)$, and/or large ``accidental" numerical factors.
The experimental analysis takes into account part of the final-state radiation,
but a consistent study of electromagnetic effects is still lacking.
In addition, in the presence of flavor-non-universal new physics, hadronic uncertainties in $R_K$
are not suppressed by $m_\mu^2/m_b^2$, but only by $(1-R_K^\text{NP})$.
}
\be
R_K = \frac{ \Gamma(B \rightarrow K \mu^+ \mu^-)}{\Gamma(B \rightarrow K e^+ e^-)}   \stackrel{\rm SM}{=}  1  + \mathcal{O}(m_{\mu}^2/m_{b}^2) \,,
\ee
was measured in the low-$q^2$ region $q^2 \in [1,6]~\text{GeV}^2$ obtaining
$R_K = 0.745^{+0.090}_{-0.074}\pm0.036$~\cite{Aaij:2014ora}, which represents a $2.6\,\sigma$ deviation from the
SM.  Other anomalies have also been observed in $B \rightarrow K^{(*)} \mu^+ \mu^-$~\cite{Aaij:2013qta,LHCb:2015dla},
and $B_s \rightarrow \phi \mu^+ \mu^-$~\cite{Aaij:2015esa}.
The exact significance of the discrepancy with the SM in the latter modes depends on the treatment of hadronic
uncertainties~\cite{1303.5794,1407.8526,Jager:2014rwa,Ciuchini:2015qxb}, but there is general consensus that sizable NP contributions
($\sim-25\%$ of the SM) to the effective operator
$(\bar s \gamma^{\alpha} P_L b) (\bar \mu \gamma_{\alpha}  \mu)$
improves the agreement with current data considerably~\cite{Descotes-Genon:2013wba,Descotes-Genon:2015uva,Altmannshofer:2014rta,Beaujean:2013soa,
Hiller:2014yaa,Ghosh:2014awa,Hurth:2016fbr}.
One key observation is that the $b\to s\mu^+\mu^-$ and $R_K$ anomalies are exactly consistent with each other
if one assumes no NP in $b\to s e^+e^-$~\cite{Ghosh:2014awa}.\\

\begin{figure}
\centering
\includegraphics[width=8.5cm,height=3.6cm]{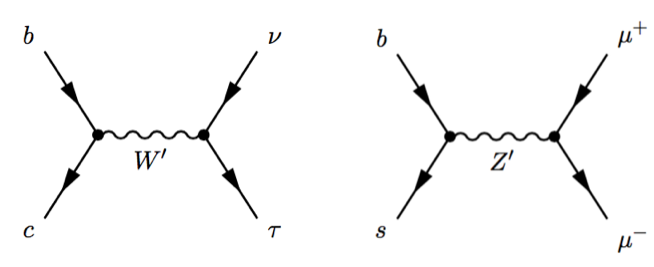} \vspace{-0.4cm}
\caption{\small \sf  Tree-level contributions to $b \rightarrow c \tau \nu$ and
$b \rightarrow s \mu^+ \mu^-$ transitions from hypothetical heavy $W', Z'$ gauge bosons.   }
\label{fig:diagramsII}
\end{figure}

A simultaneous explanation of the $b \rightarrow c \tau \nu$ and $b \rightarrow s \ell^+ \ell^-$ anomalies
has been initially discussed in Ref.~\cite{Bhattacharya:2014wla} within an effective theory (EFT) point of view,
building on the idea of Ref.~\cite{Glashow:2014iga} that non-universality in $R_K$ could be due to
NP coupling predominantly to the third generation.
This EFT approach was followed later in a series of works~\cite{Alonso:2015sja,Greljo:2015mma,Calibbi:2015kma}.  Scalar and/or vector leptoquarks as well as a $\mathrm{SU(2)}_L$ triplet of massive vector bosons coupled
predominantly to the third fermion generation were considered in these works as possible dynamical
realizations~\cite{Alonso:2015sja,Greljo:2015mma}.
The possibility of a leptoquark origin for these anomalies was subsequently explored in more
detail in Refs.~\cite{Bauer:2015knc,Fajfer:2015ycq,Barbieri:2015yvd,Bauer:2015boy,Murphy:2015kag,Deppisch:2016qqd},
making also interesting connections to other possible phenomena such as neutrino masses.

In this work we assume that the $b \rightarrow c \tau \nu$ and $b \rightarrow s \ell^+ \ell^-$ anomalies arise from
tree-level exchange of massive vector bosons (see \Fig{fig:diagramsII}). Such states could appear as heavy resonances associated to a strongly coupled dynamics~\cite{Buttazzo:2016kid}. One could also consider a scenario where these heavy vectors arise as
mediators of a perturbative short-range force.

Thus, in this work we discuss possible realizations of this idea in a minimal setting,
by extending the SM gauge group with an additional $\mathrm{SU(2)}$ factor.      
The spontaneous breaking of the enlarged gauge symmetry down to the electroweak group is supposed to
occur around the TeV scale, giving rise to heavy $W^{\prime}$ and $Z^{\prime}$ gauge bosons that can mediate
$b \rightarrow c \tau \nu$ and $b \rightarrow s \mu^+ \mu^-$ transitions as in \Fig{fig:diagramsII}.  By searching for viable models of this sort, we will see that these are restricted to be of a particular type (see \Sec{GaugeMap}). In \Sec{Model} we construct an explicit model of this type.
We identify gauge-mixing as a relevant issue for gauge models addressing
the $B$-decay anomalies. This is discussed in \Sec{g-mixing}.

\section{Interpretation of B-decay anomalies}
Measurements of decay rates as well as differential distributions in the transferred momenta and angular variables can be used to gain information about the underlying dynamics responsible for these flavor anomalies.
We make the following observations based on current data:
\begin{itemize}[leftmargin=7mm]

\item   An analysis of the $q^2 = (p_{B} - p_{D^{(*)}})^2$ differential distribution in $b \rightarrow c \tau \nu$ decays by BaBar is compatible with an electrically charged spin-1 mediator as an explanation of the $R(D^{(*)})$ excess~\cite{Lees:2013uzd}.   
 
\item Current data is compatible with the hypothesis of a universal scaling for $R(D)$ and $R(D^*)$.  This dynamical feature appears automatically for a left-handed charged current interaction~\cite{Bhattacharya:2014wla}. 

\item  Anomalies in $b \rightarrow s \mu^+ \mu^-$ transitions together with $R_K$ can be explained by the presence of a heavy neutral vector boson mediator with flavor changing couplings of the form $( \bar s \gamma^{\alpha} P_L b) Z_{\alpha}^{\prime}$, with either vectorial or left-handed coupling to muons~\cite{Descotes-Genon:2013wba,Descotes-Genon:2015uva,Altmannshofer:2014rta,Beaujean:2013soa,Ghosh:2014awa,Jager:2014rwa,Hurth:2016fbr}.

\end{itemize}

Non-abelian gauge extensions of the SM often introduce mass mixing in the gauge sector.    The latter typically appears when scalars fields responsible for the breaking of the electroweak symmetry are also charged under the extended group.  In the presence of such mixing the $W$ and $Z$ couplings receive corrections of order $v^2/M_{W^{\prime},Z^{\prime}}^2$, with additional parametric suppressions possible but model dependent.  This implies that new physics contributions from the tree-level exchange of a heavy vector boson and those due to gauge mixing effects ($W$ and $Z$ mediation) are potentially of the same size.  This is the case for $\Delta F = 1$ transitions and charged-current processes which arise at tree level in the SM.    For $\Delta F = 2$ transitions, gauge mixing effects will enter with a relative $v^2/M_{Z^{\prime}}^2$ suppression at the amplitude level compared to the tree-level exchange of a heavy $Z^{\prime}$.

Another important aspect to note is that $\Delta F = 1$ ratios probing LFU violation such as $R_{K}$ have a small sensitivity to gauge mixing effects, contrary to their charged-current counterparts, $R(D^{(*)})$ or $\Gamma(\tau \to \mu \nu\bar \nu)/\Gamma(\tau \to e \nu\bar \nu)$.   The underlying reason being that the required gauge boson couplings are already present in the SM for the case of charged-current processes.

\section{Gauge Extensions with Lepton Non-Universality}
\label{GaugeMap}

\subsection*{General considerations on model-building} 
A common explanation of universality-violating hints in the decays $b \rightarrow c \tau \nu$ and $b \rightarrow s  \ell^+ \ell^-$ poses serious challenges for model building. This is mostly because the NP mediators responsible for such processes would have to act at tree level.
Indeed, the semileptonic decays $B \rightarrow D^{(*)} \tau \nu$ are charged current processes which arise at tree level in the SM.
Since the observed deviation from the SM prediction is quite sizable, $\ord{(25\%)}$, this strongly suggests the presence of tree-level charged mediators. The same applies to the decays $B \rightarrow K^{(*)} \ell^+ \ell^-$ even though they are due, in the SM, to neutral current processes arising at one-loop level. The large deviation from the SM, again $\ord{(25\%)}$, would imply a very light mediator if the new interactions followed the SM pattern.
Such a light mediator, $\ord{(M_Z)}$, would be hard to hide from other flavor observables which are in perfect agreement with the SM as well as from direct searches for new states at high energy colliders such as the LHC.

We assume from now on that the anomalies $R_K$ and $R(D^{(*)})$ are genuine and due to new gauge bosons entering at tree level.  We are therefore looking for a non-universal gauge extension of the SM which could explain both anomalies at the same time.     We will be interested in scenarios where new physics effects in the lepton sector affect mainly the muon and tau leptons.

There are essentially two strategies to follow in constructing non-universal gauge extensions of the SM:\\[-2mm]

\noindent {\bf $\triangleright$ Non-Universality from gauge couplings (g-NU)}:\\[1mm]
via a non-universal embedding of SM fermions into a larger gauge group, or\\[-2mm]

\noindent {\bf $\triangleright$ Non-Universality from Yukawas (y-NU)}:\\[1mm]
through non-universal interactions between SM fermions and extra particles which are universally coupled to new vector bosons.\\[-2mm]

This means that, in general, non-universality is either controlled by Yukawa couplings or by gauge couplings.  Of course, one can always mix these two approaches, however we keep them separated for the sake of clarity and to gain insights based on generic considerations.
\bigskip

For simplicity and definiteness, we will focus on implementations where the gauge extensions consist of $\mathrm{SU(2)}$ and $\mathrm{U(1)}$ factors only. The minimal possibilities are denoted generically as $G(221)$ models. In addition to the source of
non-universality, $G(221)$ models can be classified according to the gauge symmetry-breaking pattern. We distinguish two broad categories:\\[-2mm]

\noindent {\bf $\triangleright$ L-Breaking Pattern (L-BP)}:\\[1mm]
For this breaking pattern the $\mathrm{U(1)}_Y$ group appears from a non-trivial breaking of the extended group:
$$\mathrm{SU(2)}_L \otimes \mathrm{SU(2)}_H \otimes \mathrm{U(1)}_H \to \EW$$

\noindent {\bf $\triangleright$ Y-Breaking Pattern (Y-BP)}:\\[1mm]
The $\mathrm{SU(2)}_L$ factor is non-trivially embedded in the extended gauge group and arises from the breaking pattern:
$$\mathrm{SU(2)}_1 \otimes \mathrm{SU(2)}_2 \otimes \mathrm{U(1)}_Y \to \EW$$

We now proceed to reviewing the viability of the different possibilities that are available in our classification.

\subsection*{Non-Universality from gauge couplings}

\noindent {\bf $\triangleright$ Breaking chain L-BP}:
A model within this scenario was already presented in Ref.~\cite{He:2012zp} to explain the $R(D^{(*)})$ anomalies.
However, just with the SM particle content one can only couple right-handed fermions to the extra gauge group,
making it unable to accommodate $R_K$.\\
  
\noindent {\bf $\triangleright$ Breaking chain Y-BP}: This model has been studied in
Refs.~\cite{Li:1981nk,Muller:1996dj,Chiang:2009kb}. In this scenario it is only possible to reproduce
the desired non-universal $Z^\prime$ and $W^\prime$ couplings to leptons if the gauge coupling hierarchy $g_2\gg g_1\sim g$
is enforced, with a single SM family coupling to $\mathrm{SU(2)}_2$. However the large $g_2$ limit has to
face constraints from rapid proton decay and perturbativity. Instanton mediated processes will, in general, 
induce proton decay when a single SM family is coupled to a non-abelian gauge group, setting a bound on the
gauge coupling: $g_2\left(M_{Z^\prime}^2\right)\lesssim 1.3-1.6$, depending on the parameters of the 
model~\cite{Morrissey:2005uza,Fuentes-Martin:2014fxa}. This bound can be circumvented by introducing
extra fermions that couple to this gauge group, such as vector-like fermions. However, even in this case perturbativity
sets an upper bound on the gauge coupling of $g_2\left(M_{Z^\prime}^2\right)<\sqrt{4\pi}\simeq3.5$. Given these limits, it is not possible
to reproduce the requested hierarchy on the lepton couplings, making this framework disfavored for the simultaneous explanation of $R_K$ and $R(D^{(*)})$.

\subsection*{Non-Universality from Yukawa couplings}

Here new vector-like (VL) fermions, charged universally under a new force to which the SM fermions are neutral, are Yukawa-coupled to the SM quarks and leptons.
The effective coupling of the SM with the new bosons is achieved via mixings with the VL fermions and is
hence controlled by the Yukawas, which can be in principle adjusted to get the desired flavor textures.
In general these mixings will also modify the SM gauge and Higgs couplings.
However one can charge the VL fermions under the gauge group in such a way that GIM protection
is enforced at the scale of the first symmetry breaking, making these deviations sufficiently small to
avoid experimental constraints.    

This translates in the two breaking patterns we are considering as follows:\\[-2mm]

\noindent {\bf $\triangleright$ Breaking chain L-BP}:
In order to obtain an effective coupling $W^{\prime\pm}$ to left-handed quarks,
it is necessary to add VL quarks which mix with the SM weak quark doublet.
The electric charge formula of this breaking chain is:
\eq{
\label{QL-BP}
Q= T_{3_L} + (T_{3_H} + H)\,,
}
where $T_{3_L}$ ($T_{3_H}$) and $H$ are respectively the isospin under $\mathrm{SU(2)}_L$ ($\mathrm{SU(2)}_H$)
and the $\mathrm{U(1)}_H$ charge. Since the SM fields are neutral under the new $\mathrm{SU(2)}_H$ interactions,
$\mathrm{U(1)}_H$ charges coincide with the standard hypercharges.
In order for two new quarks, $Q_b$ and $Q_c$ to couple to $W^{\prime\pm}$ they must belong to the same
$\mathrm{SU(2)}_H$ multiplet and their isospin must satisfy $|T_{3_H}(Q_b) -  T_{3_H}(Q_c)|=1$.
On the other hand, to preserve the GIM mechanism in the presence of the new mixings the new quarks must have
the same SM quantum numbers ($T_{3_L}$ and $Y\equiv T_{3_H} + H$) as the SM quarks with which they
mix~\cite{Langacker:1988ur}. These two requirements are in conflict with each other and so we conclude
that models of type L-BP cannot account for a unified description of $R_K$ and $R(D^{(*)})$.\\[-2mm]

\noindent  {\bf $\triangleright$ Breaking chain Y-BP}:
The product $\mathrm{SU(2)}_1 \otimes \mathrm{SU(2)}_2$ can be broken to
the diagonal $\mathrm{SU(2)}_L$ by a Higgs bi-doublet. This specific type of breaking allows for both couplings
to $W^{\prime}$ and GIM suppression. It is enough to charge SM fermions under one of the two
$\mathrm{SU(2)}\text{'}s$, say $\mathrm{SU(2)}_2$, and copy the exact same assignments for the vector-like fermions.   This is the scenario we deem more promising for the simultaneous explanation of $b\rightarrow s \ell^+ \ell^- $ and $b \rightarrow c \tau \nu$ anomalies.

\subsection*{Summary}

In summary, restricted to minimal gauge extensions we have found four broad classes of models that
lead to flavor non-universality and can potentially address the flavor anomalies.
These classes depend on the breaking pattern (L-BP or Y-BP) and the source of flavor
non-universality (g-NU or y-NU).  \Tab{tab:models} summarizes our main conclusion:
that the most promising candidates are gauge extensions where gauge couplings are
universal and non-universality arises from Yukawa couplings of the
SM fermions with a set of new vector-like fermions.

\definecolor{green1}{rgb}{0.06,0.66,0.06}
\definecolor{orange1}{rgb}{0.98,0.60,0.07}
\newcommand{\green}{{\color{green1}$\bigstar$}}
\newcommand{\orange}{{\color{orange1}\LARGE \protect\raisebox{-0.1em}{$\bullet$}}}
\newcommand{\red}{{\color{red}\small \protect\raisebox{-0.05em}{$\blacksquare$}}}
\newcommand{\one}{{\color{red}$\bigstar$}}

\begin{table}
\renewcommand{\arraystretch}{1.6}
\setlength{\tabcolsep}{7pt}
\centering
\begin{tabular}{!{\VRule[1.6pt]}c!{\VRule[1.6pt]}c!{\VRule[0.5pt]}c!{\VRule[1.6pt]}}
\toprule[1.6pt]
&L-BP & Y-BP \\
\midrule[1.4pt]
g-NU  & {\small \red~No left-handed currents}  & { \small \red~perturbativity}\\
\midrule[0.5pt]
y-NU & {\small \red~No GIM} &   \green \\
\bottomrule[1.6pt]
\end{tabular}
\caption{Summary of model building possibilities for $G(221)$ models:
source of flavor non-universality (NU) versus symmetry-breaking patterns (BP).    Blocks denote scenarios which are disfavored as an explanation of the $B$-decay anomalies while a star denotes a viable framework.      }
\label{tab:models}
\end{table}

%
\section{A Model Example}
\label{Model}

In this section we construct, as an explicit example, a model of the type y-NU/Y-BP.
We consider the gauge group
$\mathrm{SU(3)}_C \otimes \mathrm{SU(2)}_1\otimes \mathrm{SU(2)}_2\otimes \mathrm{U(1)}_Y$,
with coupling constants denoted by $(g_s,g_1,g_2,g')$ respectively. We consider also
two scalar fields transforming as:
\eq{
\phi = \left( {\bf 1} , {\bf 1} , {\bf 2} \right)_{1/2} \,, \qquad 
\Phi = \left( {\bf 1} , {\bf 2} , {\bf \bar 2} \right)_{0} \, .
}
We assume a scalar potential leading to the following vacuum-expectation values:
\begin{align}
\langle \phi \rangle = \frac{1}{\sqrt{2}}
\begin{pmatrix} 
0 \\
v \end{pmatrix},
\quad \quad 
\langle \Phi \rangle = \frac{1}{2}
\begin{pmatrix} 
u & 0 \\
0 & u \end{pmatrix},
\end{align}
with $v\simeq 246\GeV$ and $\epsilon\equiv v/u \ll 1$ (typically $u\sim\TeV$).    
The resulting symmetry-breaking pattern is given by
\begin{equation*}   
\mathrm{SU(2)}_1\otimes \mathrm{SU(2)}_2\otimes \mathrm{U(1)}_Y
\overset{u}{\rightarrow}
\mathrm{SU(2)}_L\otimes \mathrm{U(1)}_Y
\overset{v}{\rightarrow}
\mathrm{U(1)}_{\mbox{\footnotesize em}}  \, \vspace{0.3mm}
\end{equation*}
This breaking leads to a spectrum with a massless photon,
two neutral $Z,Z'$ bosons with masses
$M_Z^2 \simeq v^2 (g^2+g^{\prime 2})^2/4 $ and $M_{Z'}^2 \simeq u^2 (g_1^2+g_2^2)/4$,
and two pairs of charged bosons $W^\pm, W^{\prime\pm}$ with masses
$M_W^2 \simeq v^2 g^2/4$ and $M_{W'}^2 \simeq M_{Z'}^2$.
Here $g\equiv g_1 g_2/\sqrt{g_1^2 + g_2^2}$.

In order to source FNU from Yukawa couplings, we introduce, in addition to the SM fermion content, $\nvl$
generations of VL fermions transforming as
\begin{equation}
Q_{L,R} = \left( {\bf 3} , {\bf 2} , {\bf 1} \right)_{1/6} \, ,  \qquad    L_{L,R} = \left( {\bf 1} , {\bf 2} , {\bf 1} \right)_{-1/2}    \,.
\end{equation}
It can be shown that the requirements of {\textit{(i)}}
no $W^{\prime}/Z^{\prime}$ couplings to electrons, and {\textit{(ii)}} lepton
non-universality between $\mu$ and $\tau$, imply that $\nvl \geq 2 $.
Indeed, the first requirement introduces three non-trivial conditions on the Yukawa couplings between the chiral leptons
and the vector-like leptons which fix these completely and leave no room to satisfy the second condition.
In what follows, we take the minimal possibility and fix  $\nvl = 2 $.   The complete particle content of the model is summarized in \Tab{tab:content}.

\begin{table}[!h]
\renewcommand{\arraystretch}{1.3}
\setlength{\tabcolsep}{4.5pt}
\centering
\begin{tabular}{cccccc}
\toprule[1.4pt]   
 & generations & $\mathrm{SU(3)}_C$ & $\mathrm{SU(2)}_1$ & $\mathrm{SU(2)}_2$ & $\mathrm{U(1)}_Y$ \\
\midrule[1pt]
$\phi$ & 1 & ${\bf 1}$ & ${\bf 1}$ & ${\bf 2}$ & $1/2$ \\    \rowcolor{CGray}
$\Phi$ & 1 & ${\bf 1}$ & ${\bf 2}$ & ${\bf \bar 2}$ & $0$ \\
\midrule[1pt] 
$q_L$ & 3 & ${\bf 3}$ & ${\bf 1}$ & ${\bf 2}$ & $1/6$ \\ 
$u_R$ & 3 & ${\bf 3}$ & ${\bf 1}$ & ${\bf 1}$ & $2/3$ \\
$d_R$ & 3 & ${\bf 3}$ & ${\bf 1}$ & ${\bf 1}$ & $-1/3$ \\
$\ell_L$ & 3 & ${\bf 1}$ & ${\bf 1}$ & ${\bf 2}$ & $-1/2$ \\
$e_R$ & 3 & ${\bf 1}$ & ${\bf 1}$ & ${\bf 1}$ & $-1$ \\   \rowcolor{CGray}
$Q_{L,R}$ & $\nvl$ & ${\bf 3}$ & ${\bf 2}$ & ${\bf 1}$ & $1/6$ \\  \rowcolor{CGray}
$L_{L,R}$ & $\nvl$ & ${\bf 1}$ & ${\bf 2}$ & ${\bf 1}$ & $-1/2$ \\  
\bottomrule[1.4pt]
\end{tabular}
\caption{Particle content of the model, added fields to the SM are shown in gray. }
\label{tab:content}
\end{table}

The Dirac masses of the VL fermions are assumed to be around the symmetry breaking scale $u\sim\TeV$.   In this scenario, the couplings of $Z',W'$ bosons to right-handed SM fermions are suppressed by
$\sim m_f^2/v^2$, with $m_f$ the mass of a SM fermion, and can be neglected for the couplings we are considering.   Left-handed SM fermions have anomalous flavor-changing
couplings to $Z,W$ of $\ord(\epsilon^2)$ due to gauge mixing effects, and $\ord(1)$ couplings to $Z',W'$. The part of the
Lagrangian describing these interactions is:
\begin{widetext}
\eqa{   \label{eqmac}
\delta \L &=&
- \epsilon^2 \frac{g g_2^4}{\sqrt2 n_1^4} \Big[
(V \Delta_L^q)_{ij} W^+_\mu \bar u^i_L \gamma^\mu d^j_L 
+ (\Delta_L^\ell)_{ij} W^+_\mu \bar \nu^i_L \gamma^\mu \ell^j_L
\Big] + \text{h.c.} \nonumber\\[2mm]
&& - \frac{g g_2}{\sqrt2 g_1}
\Big[
(V \Delta_L^q)_{ij} W^{\prime +}_\mu \bar u^i_L \gamma^\mu d^j_L 
+ (\Delta_L^\ell)_{ij} W^{\prime +}_\mu \bar \nu^i_L \gamma^\mu \ell^j_L
\Big] + \text{h.c.} \nonumber\\[2mm]
&&- \epsilon^2  \frac{n_2 g_2^4}{2 n_1^4} \Big[
(V \Delta_L^q V^\dagger)_{ij} Z_\mu \bar u^i_L \gamma^\mu u^j_L
- (\Delta_L^q)_{ij} Z_\mu \bar d^i_L \gamma^\mu d^j_L
-(\Delta_L^\ell)_{ij} Z_\mu
(\bar \ell^i_L \gamma^\mu \ell^j_L-\bar \nu^i_L \gamma^\mu \nu^j_L)
\Big]\nonumber\\[2mm]
&&- \frac{g g_2}{2 g_1} \Big[
(V \Delta_L^q V^\dagger)_{ij} Z'_\mu \bar u^i_L \gamma^\mu u^j_L
- (\Delta_L^q)_{ij} Z'_\mu \bar d^i_L \gamma^\mu d^j_L
-(\Delta_L^\ell)_{ij} Z'_\mu
(\bar \ell^i_L \gamma^\mu \ell^j_L-\bar \nu^i_L \gamma^\mu \nu^j_L)
\Big]    \ ,
}
where $n_1^2\equiv g_1^2 + g_2^2$ and $n_2^2 \equiv g^2 + g^{\prime 2}$. $V$ is the CKM matrix, and
$(1-\Delta_L^{q,\ell}) \sim (\lambda\,u/M)^2$ are hermitian matrices in flavor space, with
$\lambda$ the Yukawas that couple SM and VL fermions, and $M$
the masses of the VL fermions.   The NP contributions to the relevant four-fermion operators are given by:
\eqa{
\label{mieq2}
\L_{c.c.}^{W'} &=&
-\frac{\hat g^2}{2M_{W'}^2}  (V \Delta^q_L)_{ij} (\Delta^\ell_L)_{ab}\,
(\bar u^i_L \gamma_\mu d^j_L)(\bar \ell^a_L \gamma^\mu \nu^b_L) + \text{h.c.} \,, \\[2mm]
\label{ccGM}
\L_{c.c.}^\text{GM} &=& -\frac{\hat g^2}{2M_{W'}^2}
\Big[ -(V \Delta^q_L)_{ij} \delta_{ab} - V_{ij} (\Delta^\ell_L)_{ab}\Big]\,
(\bar u^i_L \gamma_\mu d^j_L)(\bar \ell^a_L \gamma^\mu \nu^b_L) + \text{h.c.} \,, \\[2mm]
\label{mieq3}
\L_\text{FCNC}^{Z'} &=&
-\frac{\hat g^2}{4M_{W'}^2}  (\Delta^\ell_L)_{ab}\,
\Big[
(\Delta^q_L)_{ij} \, (\bar d^i_L \gamma_\mu d^j_L)
-(V \Delta^q_L V^\dagger)_{ij}\, (\bar u^i_L \gamma_\mu u^j_L)
\Big]
\Big(\bar \ell^a_L \gamma^\mu \ell^b_L -\bar \nu^a_L \gamma^\mu \nu^b_L\Big)  \,, \\[2mm]
\label{FCNCGM}
\L_\text{FCNC}^\text{GM} &=&
-\frac{\hat g^2}{4M_{W'}^2}  \delta_{ab}\,
\Big[
(\Delta^q_L)_{ij} \, (\bar d^i_L \gamma_\mu d^j_L)
-(V \Delta^q_L V^\dagger)_{ij}\, (\bar u^i_L \gamma_\mu u^j_L)
\Big]
\Big(2s_W^2\,\bar \ell^a \gamma^\mu \ell^b - \bar \ell^a_L \gamma^\mu \ell^b_L + \bar \nu^a_L \gamma^\mu \nu^b_L\Big)  \,,  \\[2mm]
\L_{\Delta F = 2}^{Z^{\prime}} &=&  - \frac{ \hat g^2 }{   8 M_{W^{\prime}}^2 }   \left[    \bigl[ ( \Delta_L^{q}   )_{ij}  (  \bar d_L^{i}  \gamma_{\mu} d_L^{j}  ) \bigr]^2  +   \bigl[   (  V \, \Delta_L^{q} \, V^{\dag}  )_{ij}  (  \bar u_L^{i}  \gamma_{\mu} u_L^{j}  )    \bigr]^2     \right] \,,
}
where $\hat g \equiv gg_2/g_1$.  Here we have separated explicitly the direct contributions from $W^{\prime}/Z^{\prime}$ exchange from those due to gauge mixing  (GM) effects.    
There are no contributions to $\L_{\Delta F=2}$ from gauge mixing at order $\epsilon^2$. This is also true for lepton-flavor non-universal ratios of
FCNCs, such as $R_K$, as can be seen from the $\delta_{ab}$ prefactor in \Eq{FCNCGM}. However, in the case
of lepton-flavor non-universal ratios in charged-current processes, such as $R(D^{(*)})$,
additional contributions from $W$-$W'$ mixing encoded in \Eq{ccGM} are present.    The effective Lagrangian for leptonic decays is given in the Appendix. \\
\end{widetext}

The conclusion of this section is that the model-building guidelines discussed in \Sec{GaugeMap}
result in models with the structure needed to address the anomalies. However, we find additional effects from gauge mixing that are typically of the same order
as the direct contributions from heavy-boson exchange.  These contributions have the potential to spoil the flavor patterns needed to explain the $B$-decay anomalies.

\section{Relevance of Gauge Mixing}
\label{g-mixing}
When present, the size of mass mixing effects in the gauge sector is intrinsically connected to the hierarchy between the electroweak scale and the scale of breaking of the extended gauge group ``$u$".  The gauge boson mass matrix receives contributions of order $u^2 W^{\prime \, 2}\,, v^2 W^2\,, v^2 W W^{\prime}$ giving rise to corrections of the electroweak gauge boson couplings of order $v^2/u^2$.  This is what occurs for example in the model presented in Sec.~\ref{Model}.    Parametrical suppressions of the mass mixing term can be engineered  in principle by considering a more involved symmetry breaking sector. 

 In this section we discuss the most relevant issues associated with gauge mixing effects when trying to account for the $B$-decay anomalies in the gauge framework of Sec.~\ref{Model}.\\[-2mm]

\noindent $\triangleright${ \bf{Bounds from $Z$ and $W$-pole observables:}}  Corrections to the electroweak gauge boson couplings to fermions are constrained at the per-mille/few percent from electroweak precision data collected at the $Z$ and $W$-pole in the LEP experiment~\cite{ALEPH:2005ab,Schael:2013ita}.    Note that when the symmetry breaking scale of the extended gauge group is of the order of the TeV scale, the natural suppression provided by the hierarchy of scales $v^2/u^2$ is typically enough to satisfy such strong bounds.

\noindent $\triangleright$ { \bf{Flavor structure/patterns:}}  The main idea behind the explanation of $B$-decay anomalies in this setting is that flavor patterns of the UV dynamics are imprinted on the low energy effective Lagrangian describing flavor transitions.   This amounts to the idea that flavor textures and/or hierarchies in the $\Delta^{q, \ell}_L$ couplings of \Eq{eqmac} can be linked to information gathered in low-energy experiments.    Gauge mixing effects however make the connection non-trivial.  The low-energy effective Lagrangian in Eqs.~\eqref{mieq2}-\eqref{FCNCGM} contains contributions due to gauge mixing which alter the flavor structure of the NP effects at low energies.   We illustrate this feature with two examples:\\[-2mm]  
  
 \textit{(i)} Fixing $\Delta^{q, \ell}_L$ to be non-vanishing only for the second and third generations can be motivated by: the strong constraints on light-quark meson systems and electrons, and, the fact that the observed $B$-decay anomalies only requires NP affecting these fermions.   One would have then vanishing NP contributions at tree level to $\Gamma(P \rightarrow \mu \nu)/\Gamma(P \rightarrow e \nu)$ (with $P = \pi, K$) from the $W^{\prime}$ exchange.  Gauge mixing effects would however introduce NP corrections to these observables via $W$-boson mediation.      The same occurs for $\mu \rightarrow e \nu \bar \nu$.\\[-2mm]

\textit{(ii)} The usual relation $\delta \C^{\mbox{\scriptsize{NP}}}_9 =  - \delta \C^{\mbox{\scriptsize{NP}}}_{10}$ associated to purely left-handed $Z'$-mediated FCNCs can receive significant corrections. Tree-level contributions from the $Z$-boson give corrections of order $v^2/M_{Z^{\prime}}^2$ to these Wilson coefficients with $|\delta \C^{\mbox{\scriptsize{NP}}}_9| \ll  |\delta \C^{\mbox{\scriptsize{NP}}}_{10}|$ due to the accidental suppression of the vectorial $Z$ coupling to leptons.\\[-2mm]

\noindent $\triangleright$ { \bf{Suppression of gauge mixing effects:}}       In the model of Sec.~\ref{Model}, such suppression could appear in certain regions of the parameter space if the scalar sector is extended by a complex scalar field with the following $\mathrm{SU(3)}_C \otimes \mathrm{SU(2)}_1 \otimes \mathrm{SU(2)}_2 \otimes \mathrm{U(1)}_Y$ quantum numbers $\phi^{\prime}(\mathbf{1},\mathbf{2},\mathbf{1})_{1/2}$.  Gauge mixing effects at low energy will depend on the ratio between the vacuum expectation values of $\phi$ and $\phi^{\prime}$, here denoted as $\tan \beta = v_{\phi}/v_{\phi^{\prime}}$.     Effectively, one can account for these changes by multiplying the right-hand side of \Eq{ccGM} and \Eq{FCNCGM} by the factor
\begin{align} \label{mixfac}
\zeta \equiv  \left( \sin^2(\beta) -  \frac{g_1^2}{g_2^2} \cos^2(\beta)  \right)    \,.
\end{align}
The effects of gauge mixing in flavor transitions are then suppressed in this scenario for $\tan(\beta) \simeq  g_1/g_2 $.  In the limit of vanishing mass mixing in the gauge sector, the structure of the low-energy effective Lagrangian relevant for flavor transitions reduces to \Eq{mieq2} and \Eq{mieq3}.

To evaluate the impact of gauge mixing effects on flavor transitions, we perform a global fit of electroweak precision and flavor data, including:    
\begin{itemize}
\item Bounds from $Z$ and $W$ pole observables, using the results provided in Ref.~\cite{Efrati:2015eaa}.           
\item  Tests of lepton universality violation in tree-level charged current processes:  $\ell \rightarrow \ell^{\prime}  \nu \bar \nu$, $\pi/K \to \ell \nu$, $\tau \to \pi/K \nu$, $K^+ \to \pi \ell \nu $, $D \to K \ell \nu$, $D_s \to \ell \nu$, $B \rightarrow D^{(*)} \ell \nu$ and $B \to X_c \ell \nu$~\cite{Agashe:2014kda,Amhis:2014hma}.
\item $|\Delta F| =1,2 $ transitions in the $b \to s$ sector receiving NP contributions at tree level in our model from the exchange of the massive neutral vector bosons.  We treat $b \to s \ell^+ \ell^-$ decays using the results of Ref.~\cite{Descotes-Genon:2015uva} and we use inputs from Ref.~\cite{Bazavov:2016nty} for $B$-meson mixing.
\item CKM inputs from a fit by the CKMfitter group with only tree-level processes~\cite{CKMgroup}, as used in Ref.~\cite{Bazavov:2016nty}.
\item Bounds from the lepton flavor violating decays $\tau \to 3 \mu$ and $Z \to \tau \mu$~\cite{Agashe:2014kda,Amhis:2014hma}.
\end{itemize}  
%

\begin{figure}
\centering
\includegraphics[width=9.5cm,height=7.2cm]{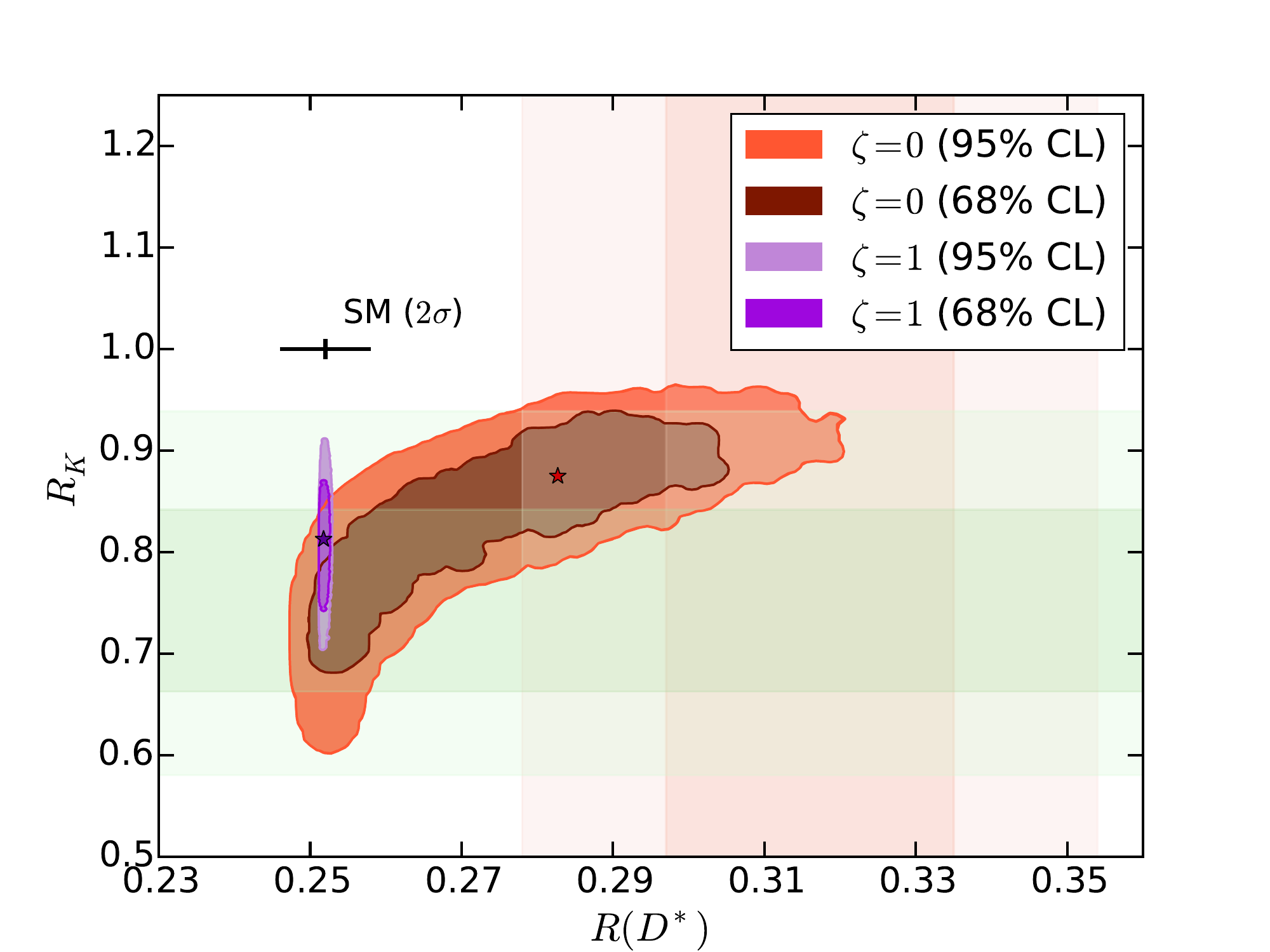} 
\caption{\small \sf   Preferred allowed regions in the $R(D^{*}) - R_K $ plane at $68\%$ and $95\%$~CL from the global fit.    Experimental values for these observables are also shown at $1\sigma$ (dark-band) and $2 \sigma$ (light-band).   The best fit points are illustrated with a star.  The SM prediction is also shown at the $2 \sigma$ level.   }
\label{fig:phenoI}
\end{figure}

We have fixed $\Delta^{q, \ell}_L$ to be non-vanishing only for the second and third generations in the fit.   Our best fit regions in the $R(D^*)-R_K$ plane are shown in \Fig{fig:phenoI} for the two benchmark values $\zeta = 0,1$ of the parameter that controls the size of gauge mixing effects.    Allowed regions at $68\%$ and $95\%$~CL in the case of  $\zeta = 1$ do not show any significant deviation from the SM in $R(D^*)$ while $R_K$ is compatible with the LHCb measurement.  In the case of vanishing gauge mixing, $\zeta = 0$, a joint explanation of the $B$-decay anomalies becomes possible as $R(D^*)$ can receive a significant enhancement compared to the SM.  Note that in our model $R(D^*)$ and $R(D)$ have the same NP scaling since the $W^{\prime}$ has left-handed couplings to fermions.

\section{Conclusions}
New Physics models with vector-boson triplets are potential candidates to accommodate the
anomalies in $b\to c\ell\nu$ and $b\to s\ell\ell$. It has been shown that effective models
with generic contributions to dimension-six operators, as well as more concrete dynamical models,
can fit the anomalies while satisfying stringent constraints from flavor-universality.
The question is whether one can build concrete gauge models of this sort.  

We find that minimal gauge extensions of the SM leading to heavy gauge-boson triplets must be of
a very particular type in order to address the $B$-decay anomalies and at the same time
satisfy other constraints such as perturbativity, GIM suppression, or proton decay. We identify
a viable class of gauge extensions in which the $\mathrm{SU(2)}_L$ factor is embedded non-trivially in the extended gauge group.
Flavor non-universality is sourced by Yukawa couplings to additional matter fields, such as VL fermions.
 
We have built a concrete model and checked that it
reproduces the correct patterns for the dimension-six operators related to the anomalies. We have identified the
issue of gauge mixing as a relevant obstacle towards a
joint explanation of the $B$-decay anomalies in the con-
text of gauge extensions of the SM. The impact of these
mixing effects on the explanation of the $B$-decay anomalies is illustrated in \Fig{fig:phenoI}. 

We find that a joint explanation of the $B$-decay anomalies is only possible within the proposed gauge framework when gauge mixing effects are suppressed. This is
achieved via a non-trivial tuning between parameters of
the scalar and gauge sector.

If lepton-flavor non-universality is established through more precise
measurements in $B$ decays and a more thorough examination of theoretical uncertainties, this could be
the first indication of an extended gauge symmetry, such as the one presented here.

\section*{Acknowledgements}
\linespread{1}\selectfont
{\small
We thank Thorsten Feldmann, S\'ebastien Descotes-Genon and Admir Greljo for discussions.   We also thank Adam Falkowski for providing numerical results used in our analysis of electroweak precision data.
S.M.B. thanks LMU for kind hospitality, and acknowledges support of the MIUR grant for 
the Research Projects of National Interest PRIN 2012 No.2012CP-PYP7 Astroparticle Physics, of INFN I.S. TASP2014, and of  MultiDark CSD2009-00064.
The work of A.C. is supported by the Alexander von Humboldt Foundation.   
The work of J.F. is supported in part by the Spanish Government, ERDF from the EU Commission and the Spanish  Centro  de  Excelencia Severo  Ochoa  Programme [Grants~No.~FPA2011-23778, FPA2014-53631-C2-1-P, PROMETEOII/2013/007, SEV-2014-0398].  J.F.  also acknowledges VLC-CAMPUS for an ``Atracci\'{o} del Talent"  scholarship.
A.V. acknowledges financial support from the ``Juan de la Cierva'' program
(27-13-463B-731) funded by the Spanish MINECO as well as from the
Spanish grants FPA2014-58183-P, Multidark CSD2009-00064,
SEV-2014-0398 and PROMETEOII/2014/084
(Generalitat Valenciana).   
J.V. is funded by the DFG within research unit FOR 1873 (QFET).
A.C. and J.V. are grateful to the Munich Institute for Astro- and Particle Physics (MIAPP) of the DFG cluster of excellence ``Origin and Structure of the Universe" for support while part of this work was done.

}

\begin{appendix}

\section{Effective Lagrangian for leptonic decays}
The effective Lagrangian for lepton flavor violating decays $\ell_b \to \ell_a \bar \ell_c \ell_c$ is given by
\eqa{
\label{mieq20}
\L_{\mbox{\scriptsize{LFV}}} &=&  - \frac{\hat g^2}{4 M_{W^{\prime}}^2}  (\Delta_L^{\ell})_{ab} (  \Delta_L^{\ell} )_{cc}  \left(   \bar \ell_L^{a} \, \gamma_{\mu} \, \ell_L^b  \right) \left(   \bar \ell_L^{c} \, \gamma^{\mu} \, \ell_L^c \right)   \nonumber \\
&-& \frac{\hat g^2}{4 M_{W^{\prime}}^2} \zeta  (2 s_W^2 - 1) (\Delta_L^{\ell})_{ab} \left(   \bar \ell_L^{a} \, \gamma_{\mu} \, \ell_L^b  \right) \left(   \bar \ell_L^{c} \, \gamma^{\mu} \, \ell_L^c \right) \nonumber \\
&-& \frac{\hat g^2}{2 M_{W^{\prime}}^2}  \zeta  s_W^2  (\Delta_L^{\ell})_{ab} \left(   \bar \ell_L^{a} \, \gamma_{\mu} \, \ell_L^b  \right) \left(   \bar \ell_R^{c} \, \gamma^{\mu} \, \ell_R^c \right) \,. 
}
Here we have included the effect of the doublet $\phi^{\prime}$ on the gauge mixing through the parameter $\zeta$, see~\Eq{mixfac}.   For leptonic decays conserving lepton flavor $\ell \rightarrow \ell^{\prime}  \bar \nu_{\ell^{\prime}} \nu_{\ell}$ the effective Lagrangian is 
\eqa{
\label{mieq30}
\L_{\mbox{\scriptsize{LFNU}}} &=&  - \frac{\hat g^2}{4 M_{W^{\prime}}^2}  \Bigl\{  \left[    2   (  \Delta_L^{\ell} )_{ad} ( \Delta_L^{\ell} )_{cb}  -   (  \Delta_L^{\ell} )_{ab} ( \Delta_L^{\ell} )_{cd}  \right]   \nonumber \\
&-&  2 \zeta \, \left[  \delta_{ad}  (   \Delta_L^{\ell} )_{cb} + \delta_{cb}  (\Delta_L^{\ell})_{ad}     \right]   \Bigl\}    \left(    \bar \ell_L^{a}  \gamma_{\mu} \ell_L^{b}     \right)  \left(  \bar \nu_L^c   \gamma^{\mu}  \nu_L^{d}       \right)  .  \nonumber \\
}

\end{appendix}

\end{document}